\RequirePackage[switch,columnwise]{lineno}
\documentclass[aps,prd,twocolumn, superscriptaddress, showkeys]{revtex4-1}  
\usepackage{graphicx}  
\usepackage{dcolumn}   
\usepackage{bm}        
\usepackage{amssymb}   
\usepackage{graphicx, epsfig, subfigure}
\graphicspath{{figures/}}

\usepackage{color}

\begin{document}

    \title{ Measurements of the transverse-momentum-dependent cross sections of $J/\psi$ production at mid-rapidity in proton+proton collisions at $\sqrt{s} =$ 510 and 500 GeV with the STAR detector } 

\affiliation{Abilene Christian University, Abilene, Texas   79699}
\affiliation{AGH University of Science and Technology, FPACS, Cracow 30-059, Poland}
\affiliation{Alikhanov Institute for Theoretical and Experimental Physics, Moscow 117218, Russia}
\affiliation{Argonne National Laboratory, Argonne, Illinois 60439}
\affiliation{Brookhaven National Laboratory, Upton, New York 11973}
\affiliation{University of California, Berkeley, California 94720}
\affiliation{University of California, Davis, California 95616}
\affiliation{University of California, Los Angeles, California 90095}
\affiliation{University of California, Riverside, California 92521}
\affiliation{Central China Normal University, Wuhan, Hubei 430079 }
\affiliation{University of Illinois at Chicago, Chicago, Illinois 60607}
\affiliation{Creighton University, Omaha, Nebraska 68178}
\affiliation{Czech Technical University in Prague, FNSPE, Prague 115 19, Czech Republic}
\affiliation{Technische Universit\"at Darmstadt, Darmstadt 64289, Germany}
\affiliation{E\"otv\"os Lor\'and University, Budapest, Hungary H-1117}
\affiliation{Frankfurt Institute for Advanced Studies FIAS, Frankfurt 60438, Germany}
\affiliation{Fudan University, Shanghai, 200433 }
\affiliation{University of Heidelberg, Heidelberg 69120, Germany }
\affiliation{University of Houston, Houston, Texas 77204}
\affiliation{Huzhou University, Huzhou, Zhejiang  313000}
\affiliation{Indiana University, Bloomington, Indiana 47408}
\affiliation{Institute of Physics, Bhubaneswar 751005, India}
\affiliation{University of Jammu, Jammu 180001, India}
\affiliation{Joint Institute for Nuclear Research, Dubna 141 980, Russia}
\affiliation{Kent State University, Kent, Ohio 44242}
\affiliation{University of Kentucky, Lexington, Kentucky 40506-0055}
\affiliation{Lawrence Berkeley National Laboratory, Berkeley, California 94720}
\affiliation{Lehigh University, Bethlehem, Pennsylvania 18015}
\affiliation{Max-Planck-Institut f\"ur Physik, Munich 80805, Germany}
\affiliation{Michigan State University, East Lansing, Michigan 48824}
\affiliation{National Research Nuclear University MEPhI, Moscow 115409, Russia}
\affiliation{National Institute of Science Education and Research, HBNI, Jatni 752050, India}
\affiliation{National Cheng Kung University, Tainan 70101 }
\affiliation{Nuclear Physics Institute of the CAS, Rez 250 68, Czech Republic}
\affiliation{Ohio State University, Columbus, Ohio 43210}
\affiliation{Institute of Nuclear Physics PAN, Cracow 31-342, Poland}
\affiliation{Panjab University, Chandigarh 160014, India}
\affiliation{Pennsylvania State University, University Park, Pennsylvania 16802}
\affiliation{NRC "Kurchatov Institute", Institute of High Energy Physics, Protvino 142281, Russia}
\affiliation{Purdue University, West Lafayette, Indiana 47907}
\affiliation{Pusan National University, Pusan 46241, Korea}
\affiliation{Rice University, Houston, Texas 77251}
\affiliation{Rutgers University, Piscataway, New Jersey 08854}
\affiliation{Universidade de S\~ao Paulo, S\~ao Paulo, Brazil 05314-970}
\affiliation{University of Science and Technology of China, Hefei, Anhui 230026}
\affiliation{Shandong University, Qingdao, Shandong 266237}
\affiliation{Shanghai Institute of Applied Physics, Chinese Academy of Sciences, Shanghai 201800}
\affiliation{Southern Connecticut State University, New Haven, Connecticut 06515}
\affiliation{State University of New York, Stony Brook, New York 11794}
\affiliation{Temple University, Philadelphia, Pennsylvania 19122}
\affiliation{Texas A\&M University, College Station, Texas 77843}
\affiliation{University of Texas, Austin, Texas 78712}
\affiliation{Tsinghua University, Beijing 100084}
\affiliation{University of Tsukuba, Tsukuba, Ibaraki 305-8571, Japan}
\affiliation{United States Naval Academy, Annapolis, Maryland 21402}
\affiliation{Valparaiso University, Valparaiso, Indiana 46383}
\affiliation{Variable Energy Cyclotron Centre, Kolkata 700064, India}
\affiliation{Warsaw University of Technology, Warsaw 00-661, Poland}
\affiliation{Wayne State University, Detroit, Michigan 48201}
\affiliation{Yale University, New Haven, Connecticut 06520}

\author{J.~Adam}\affiliation{Creighton University, Omaha, Nebraska 68178}
\author{L.~Adamczyk}\affiliation{AGH University of Science and Technology, FPACS, Cracow 30-059, Poland}
\author{J.~R.~Adams}\affiliation{Ohio State University, Columbus, Ohio 43210}
\author{J.~K.~Adkins}\affiliation{University of Kentucky, Lexington, Kentucky 40506-0055}
\author{G.~Agakishiev}\affiliation{Joint Institute for Nuclear Research, Dubna 141 980, Russia}
\author{M.~M.~Aggarwal}\affiliation{Panjab University, Chandigarh 160014, India}
\author{Z.~Ahammed}\affiliation{Variable Energy Cyclotron Centre, Kolkata 700064, India}
\author{I.~Alekseev}\affiliation{Alikhanov Institute for Theoretical and Experimental Physics, Moscow 117218, Russia}\affiliation{National Research Nuclear University MEPhI, Moscow 115409, Russia}
\author{D.~M.~Anderson}\affiliation{Texas A\&M University, College Station, Texas 77843}
\author{R.~Aoyama}\affiliation{University of Tsukuba, Tsukuba, Ibaraki 305-8571, Japan}
% ...

\collaboration{STAR Collaboration}\noaffiliation

\date{\today}

\begin{abstract}
    We present measurements of the differential cross sections of  inclusive $J/\psi$ meson production as a function of transverse momentum ($p_{T}^{J/\psi}$) using the $\mu^{+}\mu^{-}$ and $e^{+}e^{-}$ decay channels in proton+proton collisions at center-of-mass energies of 510 and 500 GeV, respectively, recorded by the STAR detector at the Relativistic Heavy Ion Collider.  
    The measurement from the $\mu^{+}\mu^{-}$ channel is for 0 $< p_{T}^{J/\psi} <$ 9 GeV/$c$ and rapidity range $|y^{J/\psi}| < $ 0.4, and that from the $e^{+}e^{-}$ channel is for 4 $< p_{T}^{J/\psi} <$ 20 GeV/$c$ and $|y^{J/\psi}| < $ 1.0.  
    The $\psi(2S)$ to $J/\psi$ ratio is also measured for 4 $< p_{T}^{\rm meson} <$ 12 GeV/$c$ through the $e^{+}e^{-}$ decay channel. 
    Model calculations, which incorporate different approaches toward the $J/\psi$ production mechanism, are compared with experimental results and show reasonable agreement within uncertainties. 
 
\end{abstract}


\maketitle

\section{Introduction}
The $J/\psi$ meson is a bound state of charm and anti-charm quarks ($c\bar{c}$) which was discovered several decades ago~\cite{jpsi_orig}. 
In hadronic collisions at energies reached at the Relativistic Heavy Ion Collider (RHIC), $J/\psi$ are primarily produced via inelastic scattering by two gluons into charm and anti-charm quarks, followed by hadronization of the $c\bar{c}$ pair~\cite{jpsi_1, jpsi_2}. 
Studying the $J/\psi$ production provides valuable knowledge for the understanding of Quantum Chromodynamics (QCD) in both perturbative and non-perturbative regimes.
The production of the $c\bar{c}$ pair can be calculated using the perturbative approach, however, the evolution of a $c\bar{c}$ pair into a $J/\psi$ meson is non-perturbative, and the theoretical description remains a challenge. 
Different theoretical approaches have been proposed to describe the $J/\psi$ production mechanism~\cite{csm,nrqcd,cem,icem,cgc}. 
However, these descriptions have difficulties in explaining the experimental results of production cross section and polarization simultaneously. 
Therefore, precise measurements of the $J/\psi$ cross section in elementary collisions at different collision energies are essential for investigating the $J/\psi$ production mechanism. 
Moreover, as an important probe of the hot and dense medium, known as quark-gluon plasma (QGP), it is necessary to have a good understanding of the $J/\psi$ production mechanism in elementary collisions in order to help understand the modification to its production in heavy-ion collisions, which has been proposed and widely pursued to study the properties of QGP~\cite{jpsi_supp}.  

There are three notable models for $J/\psi$ production which differ mainly in the description of the non-perturbative process. 
These are the Color Singlet Model (CSM)~\cite{csm}, Non-Relativistic QCD formalism (NRQCD)~\cite{nrqcd} and the Color Evaporation Model (CEM)~\cite{cem}.
In the CSM, it is assumed that the hadronization process does not change the quantum numbers of the $c\bar{c}$ pair. 
The initially produced $c\bar{c}$ can then bind to a given charmonium state only if it is created in a color-singlet state with matching angular-momentum quantum numbers.
The Next-to-Next-to-Leading-Order CSM (NNLO$^{\star}$ CSM) has been tested for the $S$-wave quarkonium states in the Tevatron and LHC data. 
However, this model is not able to calculate the full NNLO contribution, or provide the predictions for the $P$-wave states, due to limitations of accuracy in the NLO calculation~\cite{csm_nnlo}.
Therefore, it is expected to underestimate the production for the quarkonium states which have significant contributions from the decays of excited states, known as feed-down contributions~\cite{atlas_upsi, cms_upsi}. 
In the NRQCD approach, the charmonium can be produced from both the CS state and a color-octet (CO) state. 
The color neutralization of the CO state is achieved by radiating soft gluons during the hadronization process. 
In the CEM, the produced $c\bar{c}$ pair is assumed to evolve into a $J/\psi$ with a certain probability if its invariant mass is below the threshold for producing a $D\overline{D}$ pair.
In this model, spin is always summed over which prevents it from predicting the $J/\psi$ polarization.
A recent improvement to the CEM (Improved CEM (ICEM)~\cite{icem}) overcomes this issue by sorting out different spin states and is able to predict the polarization of the quarkonium states. 
In the low transverse momentum ($p_T$) range of the charmonium, the $c\bar{c}$ cross section becomes difficult to calculate at collider center-of-mass energies since the dynamics are sensitive to the large logarithms of small Bjorken $x$.
A newly developed Color Glass Condensate (CGC) effective theory of small-$x$ QCD provides a viable path towards calculating the $J/\psi$ cross section at low $p_{T}$ ($p_{T}$ $<$ $\sim$$M$, where $M$ is the quarkonium mass) by combining the CGC effective theory with the NRQCD formalism~\cite{cgc}. 

This paper presents the measurements of the $J/\psi$ production cross sections covering a wide $p_{T}$ range from 0 to 20 GeV/$c$ in proton+proton collisions at center-of-mass energies of 510 and 500 GeV at RHIC. 
These cross sections are measured in two decay channels, which include the $\mu^{+}\mu^{-}$ channel, for 0 $< p_{T}^{J/\psi} <$ 9 GeV/$c$ and $J/\psi$ rapidity ($|y^{J/\psi}|$) $<$ 0.4, and the $e^{+}e^{-}$ channel, for 4 $< p_{T}^{J/\psi} <$ 20 GeV/$c$ and $|y^{J/\psi}| <$ 1.0, respectively. 
The measured cross sections contain the direct production of $J/\psi$, contributions from excited charmonium states, and from decays of bottom-flavored hadrons.
The first two are often categorized as prompt $J/\psi$ as they are produced at the collision vertex and cannot be experimentally separated. 
The last one is often called non-prompt $J/\psi$, while the detector setup used in this analysis cannot experimentally distinguish it from prompt $J/\psi$.
The feed-down contribution to the $J/\psi$ production is an additional complication in understanding the $J/\psi$ production mechanism as nearly 30 $-$ 40$\%$ of the inclusive $J/\psi$ yields come from the decay of excited charmonium states~\cite{jpsi_feed_1, jpsi_feed_2}. 
Many experiments have already presented the results of heavy quarkonium production in electron+positron, hadron+hadron, and heavy-ion collisions~\cite{jpsi_200, jpsi_ref}. 
The latest measurements from the LHC~\cite{jpsi_atlas, jpsi_cms, jpsi_lhcb} probe the high $p_T$ production cross sections in proton+proton collisions with center-of-mass energies of 7, 8, and 13 TeV. 
The large kinematic range of the $J/\psi$ measurement at the highest beam energies at RHIC (510 and 500 GeV) provides valuable insights to the $J/\psi$ production mechanism. 
Additionally, the $\psi(2S)$ to $J/\psi$ ratio is measured in the $e^{+}e^{-}$ decay channel in the $p_{T}$ range of 4 $< p_{T}^{\rm meson} <$ 12 GeV/$c$. This measurement could help constrain the feed-down contribution to the $J/\psi$ from the excited charmonium states. 

The paper is organized as follows: the STAR detector will be discussed in section II, and the analyses of the $\mu^{+}\mu^{-}$ and $e^{+}e^{-}$ decay channels will be described in detail in section III and IV, respectively. 
The results from these two different channels will be presented and compared to different theoretical models in section V. 
Finally, conclusions will be given in section VI.

\section{The STAR Detector}
The STAR detector is optimized for high energy nuclear physics. 
It has excellent particle identification capability and a large acceptance at mid-rapidity.
The heart of the STAR detector is the Time Projection Chamber (TPC).
The TPC is the primary tracking detector for charged particles and provides particle identification via measurements of their ionization energy losses ($dE/dx$)~\cite{tpc}.  
It covers the full azimuthal range ($0 \leq \phi < 2 \pi$) and a large pseudorapidity range ($|\eta| < 1$). 
The $p_T$ of charged particles are measured from the curvature of their trajectories in the 0.5 Tesla solenoidal field~\cite{magnet}.
There are 30 iron bars, known as ``backlegs'' outside the coil to provide the return flux path for the magnetic field. 
These are 61 cm thick at a radius of 363 cm corresponding to about 5 interaction lengths.
These backlegs play an essential role in enhancing the muon purity by absorbing the background hadrons from collisions. 
The hadron rejection rate is about 99\% as shown in the simulation study~\cite{mtd}.
The Muon Telescope Detector (MTD) is a fast detector which uses Multi-gap Resistive Plate Chamber technology to record signals, also referred to as ``hits'', generated by charged particles traversing it. 
It provides single-muon and dimuon triggers depending on the number of hits within a predefined online timing window. 
The MTD modules are installed at a radius of about 403 cm, and the full MTD detector covers about 45\% in azimuth within $|\eta| <$ 0.5~\cite{mtd}.
The timing resolution of the MTD is $\sim$100 ps and the spatial resolutions are $\sim$1-2 cm in both $r\phi$ and $z$ directions as demonstrated in the cosmic-ray data~\cite{mtd_3}. 
The data used in this analysis were taken during the run in which the MTD detector was 63\% completed.
The Barrel Electromagnetic Calorimeter (BEMC) is a lead-scintillator sampling calorimeter with 23 radiation lengths~\cite{bemc}. 
The BEMC, being a thick absorber, is dedicated to measuring energies of particles with electromagnetic interactions, such as electrons and positrons. 
The BEMC is physically segmented into a total of 4800 towers with a granularity of $0.05 \times 0.05$ in $\Delta\phi \times \Delta\eta$. 
The energy deposited in the towers is used as a trigger to record rare events.  
The Vertex Position Detectors (VPD)~\cite{vpd} and the Beam Beam Counters (BBC)~\cite{bbc} are scintillator-based detectors located on both sides of the main detector, and they cover pseudorapidity ranges from 4.4 to 4.9 and 2 to 5, respectively.


\section{Measurement of $J/\psi \rightarrow \mu^{+}\mu^{-}$ Signal}

\subsection{Data and Monte Carlo}
Data for the $\mu^{+}\mu^{-}$ channel in this analysis were collected by the STAR detector during the 2013 RHIC proton+proton run at a collision energy of 510 GeV. 
The corresponding integrated luminosity sampled by the MTD dimuon trigger, which requires at least two coincidence hits on the MTD, as well as signals in the VPD and the two BBCs, within the bunch crossing is 22.0 $pb^{-1}$.
Events used in the analysis are required to have a valid reconstructed vertex with at least two tracks that are associated with corresponding MTD hits.

A Monte Carlo (MC) simulation sample was generated by a single-particle generator with flat distributions in $p_T$, $\phi$ and $y$ for the $J/\psi$ signal. 
These simulated signals were passed through a full GEANT3~\cite{geant3} STAR detector simulation and then ``embedded'' into real data events. 
These embedded events were reconstructed using the same reconstruction procedure used for real data.
The kinematic distributions of the embedded $J/\psi$ were weighted by the $p_T$ spectrum of $J/\psi$ in proton+proton collisions at a collision energy of 510 GeV, determined via interpolation through a global fit of world-wide measurements of $J/\psi$ cross sections~\cite{jpsi_pt}.
Due to the systematic uncertainties on various distortion corrections for the TPC, the $p_T$ resolution of the reconstructed muon in MC was retuned to match the reconstructed $J/\psi$ signal mass shape in data.

\subsection{Muon candidate selection}
\begin{figure}[!htbp]
  \begin{center}
      \includegraphics[width=0.42\textwidth]{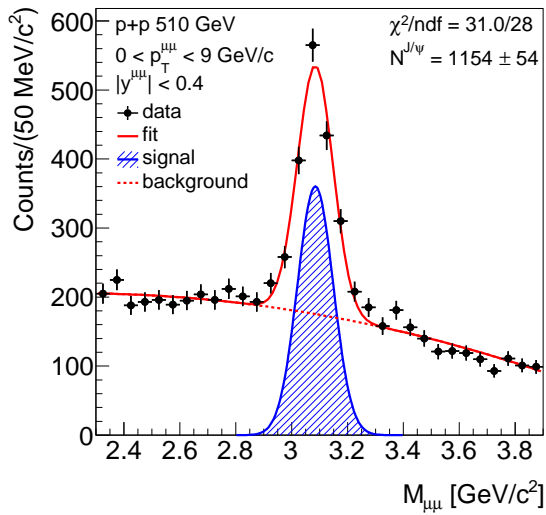}
      \end{center}
    \caption{  
   The $\mu^{+}\mu^{-}$ invariant mass spectrum in proton+proton collisions at $\sqrt{s}$ = 510 GeV. The red solid line depicts a fit using a Gaussian function (blue line) for $J/\psi$ signal and a second-order polynomial function (red dashed line) for background.
  \label{fig:dimu_mass}}
\end{figure}

Muon candidates for reconstructing the $J/\psi$ signal must satisfy the following selection criteria:  
$p_T^{\mu}$ is greater than 1.3 GeV/$c$ to ensure the track can reach the MTD detector; 
the pseudorapidity of the track is within the MTD acceptance, $|\eta^{\mu}| < 0.5$; 
the distance of closest approach (DCA) to the collision vertex must be less than 3 cm to suppress background tracks from pile-up events and secondary decay vertices;
the number of TPC clusters used in track reconstruction is more than 15 (the maximum possible is 45) to ensure good momentum resolution; 
the number of TPC clusters used for the $dE/dx$ measurement should be more than 10 to have good $dE/dx$ resolution;
the ratio of the number of TPC clusters used over the number of possible clusters is at least 0.52 to avoid double counting for the same tracks from track splitting. 
Tracks are propagated from the interaction vertex to the MTD and required to match the MTD hits geometrically which fired the trigger. 
In addition, the muon candidates were selected by an advanced muon identification method called the Likelihood Ratio method which is described in Ref.~\cite{muid}. 
The rapidity of the $\mu^{+}\mu^{-}$ pairs should be smaller than 0.4 to reduce the edge effect from the $J/\psi$ kinematic acceptance  which will be described in the next section. 
Figure~\ref{fig:dimu_mass} shows the invariant mass spectrum of the $\mu^{+}\mu^{-}$ pairs with the selection criteria described above applied to both candidate daughters. 
This can be well described by a single Gaussian as signal plus second-order polynomial function as background. 
A total of 1154 $\pm$ 54 final $J/\psi$ candidates are observed within the kinematic phase space of 0 $ < p_T^{J/\psi} < $ 9 GeV/$c$ and $|y^{J/\psi}| < $ 0.4. 

\subsection{Acceptance and efficiency}
The differential production cross section multiplied by the $J/\psi \to \mu^+\mu^-$ branching ratio (BR), ($5.961 \pm 0.033$)\%~\cite{pdg_jpsi}, is given by
\begin{eqnarray}
    {\rm BR}\times\frac{d^2\sigma}{2\pi p_T dp_T dy} = \frac{N_{J/\psi\to\mu^{+}\mu^{-}}^{\rm corr.}}{(2\pi p_T) \cdot \int \mathcal{L} dt \cdot \Delta p_T \cdot \Delta y}, 
\label{eq:xsec}
\end{eqnarray}
where $N_{J/\psi\to\mu^{+}\mu^{-}}^{\rm corr.}$ is the efficiency-corrected number of $J/\psi$ candidates. 
$\int \mathcal{L} dt $ is the corresponding integrated luminosity. 
$\Delta p_T$ and $\Delta y$ are the corresponding bin widths in $p_T$ and $y$ of the $\mu^{+}\mu^{-}$ pairs, respectively.  
For each $\mu^{+}\mu^{-}$ candidate, a weighting factor ($w_i$) is multiplied to correct for the detector acceptance ($\mathcal{A}$) and the total reconstruction efficiency ($\varepsilon_{\rm reco.}$), to obtain $N_{J/\psi\to\mu^{+}\mu^{-}}^{\rm corr.} = \sum_{i=1}^{N_{J/\psi}} w_i$, where $w_i^{-1} = \mathcal{A} \times \varepsilon_{\rm reco.}$ and $i$ indicates the $i$th candidate. 
This minimizes the potential bias from the efficiency correction due to the gaps between the MTD modules and restricted pseudorapidity coverage.

\begin{figure}[!htbp]
  \begin{center}
      \includegraphics[width=0.42\textwidth]{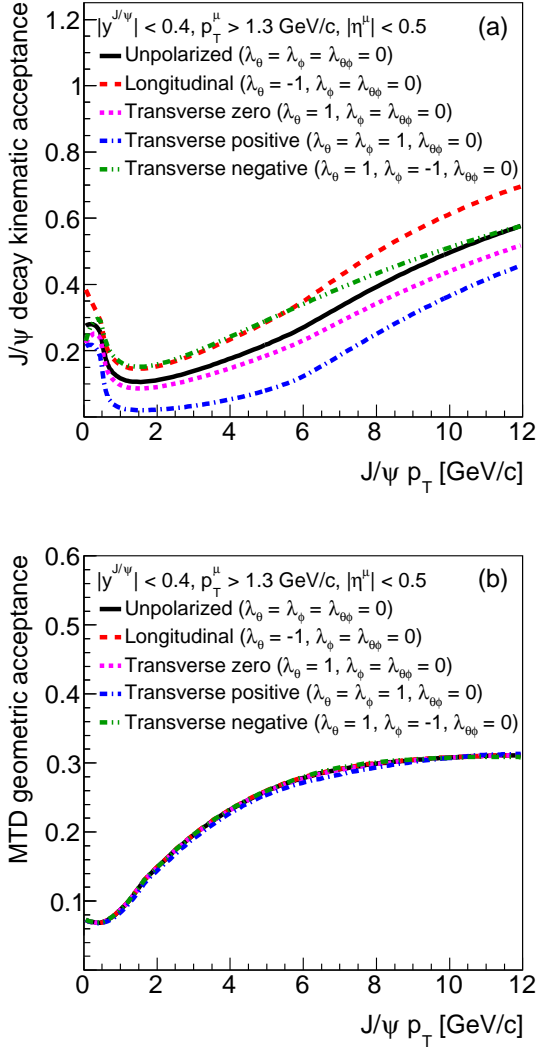}
  \end{center}
    \caption{ (a) The $J/\psi$ decay kinematic acceptance and (b) the MTD geometric acceptance as a function of $J/\psi$ $p_T$ with different polarization assumptions in the $J/\psi \to \mu^{+} \mu^{-}$ analysis. The black solid line is the unpolarized, the red dashed line is longitudinally polarized, the pink dotted line is zero transversely polarized, the blue dashed-dotted line is positively transversely polarized, and the green dashed-dotted-dotted line is negatively transversely polarized.
  \label{fig:tot_accp}}
\end{figure}

The detector acceptance, $\mathcal{A}$, is the probability of detecting muons having certain kinematics, namely $p_T^{\mu} >$ 1.3 GeV/$c$ and $|\eta| < 0.5$, from the $J/\psi$ decay within the detector fiducial volume. 
The $\mathcal{A}$ can be factorized into the $J/\psi$ decay kinematic acceptance and the MTD geometric acceptance, $\mathcal{A}$ = $\mathcal{A}_{J/\psi}$ $\times$ $\mathcal{A}_{\rm MTD}$. 
The acceptance can be determined with the muon angular distribution calculated in the $J/\psi$ rest frame by the following formula~\cite{pol}:   
\begin{eqnarray}
    \frac{d^2N}{d\cos\theta^{\star}d\phi^{\star}} \propto &1& + \lambda_{\theta}\cos^2\theta^{\star}\ + \lambda_{\phi}\sin^2\theta^{\star}\cos2\phi^{\star}  \nonumber \\
                                                          &+& \lambda_{\theta\phi}\sin2\theta^{\star}\cos\phi^{\star},  
\label{eq:eff}
\end{eqnarray} 
where $\theta^{\star}$ is the polar angle between the $\mu^{+}$ momentum in the $J/\psi$ rest frame and the direction of the $J/\psi$ momentum in the laboratory frame; $\phi^{\star}$ is the azimuthal angle between the $J/\psi$ production plane (defined in the $J/\psi$ rest frame by the momenta of the incoming protons) and the $J/\psi$ decay plane in the lab frame; and $\lambda_i$ are the parameters for different polarization configurations.  
Similar to the analyses carried out by other experiments~\cite{jpsi_atlas, atlas_upsi}, five extreme configurations are considered to cover the polarization phase space: unpolarized, $\lambda_{\theta} = \lambda_{\phi} = \lambda_{\theta\phi} = 0$; longitudinaly polarized, $\lambda_{\theta} = -1, \lambda_{\phi} = \lambda_{\theta\phi} = 0$; zero transversely polarized, $\lambda_{\theta} = +1, \lambda_{\phi} = \lambda_{\theta\phi} = 0$; positively transversely polarized, $\lambda_{\theta} = +1, \lambda_{\phi} = +1, \lambda_{\theta\phi} = 0$; and negatively transversely polarized, $\lambda_{\theta} = +1, \lambda_{\phi} = -1, \lambda_{\theta\phi} = 0$. 
Figure~\ref{fig:tot_accp} shows the $J/\psi$ decay kinematics acceptance  and the MTD geometric acceptance as a function of $J/\psi$ $p_T$ with different polarization assumptions, respectively. 
There is a significant difference in the $J/\psi$ decay kinematic acceptance for different polarization assumptions at $p_T$ around 2 GeV/$c$, and the fractional difference becomes smaller at higher $p_T$. 
On the other hand, the MTD geometric acceptance is almost independent of the $J/\psi$ polarization configuration.

\begin{figure}[!htbp]
  \begin{center}
      \includegraphics[width=0.42\textwidth]{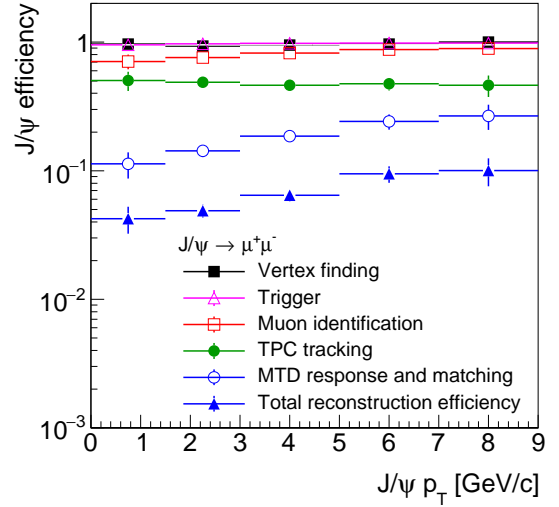}
      \end{center}
  \caption{  
      The total $J/\psi$ efficiency as a function of $J/\psi$ $p_T$ in the $J/\psi \to \mu^{+} \mu^{-}$ analysis. Individual contributions are also shown. The green solid circles are the TPC tracking efficiency; the magenta open triangles are the trigger efficiency; the blue open circles are the MTD efficiency including response and matching; the red open squares are the muon identification efficiency; and the blue solid triangles are the total reconstruction efficiency.  
  \label{fig:tot_eff}}
\end{figure}

The total $J/\psi$ reconstruction efficiency, $\varepsilon_{\rm reco.}$, includes the VPD requirement, the TPC tracking efficiency, the vertex-finding efficiency, the dimuon triggering, the MTD in-situ response and matching, and muon-identification efficiencies, as shown in the following: 
\begin{eqnarray}
    \varepsilon_{\rm reco.} = && \varepsilon_{\rm VPD} \times \varepsilon_{\rm vtx} \times \varepsilon_{\rm trig.} \times (\varepsilon_{\rm trk}^{1} \cdot \varepsilon_{\rm MTD}^{1} \cdot \varepsilon_{\rm ID}^{1}) \nonumber \\ 
                         && \times  (\varepsilon_{\rm trk}^{2} \cdot \varepsilon_{\rm MTD}^{2} \cdot \varepsilon_{\rm ID}^{2}),  
\label{eq_eff}
\end{eqnarray}
where the superscripts, 1 and 2, indicate the first and second muon from a $J/\psi$ candidate.
The VPD efficiency ($\varepsilon_{\rm VPD}$) is obtained from the zero-bias MC sample. 
The TPC tracking efficiency ($\varepsilon_{\rm trk}$) is calculated from the $J/\psi \to \mu^{+} \mu^{-}$ MC sample. 
The vertex finding efficiency ($\varepsilon_{\rm vtx}$) is obtained from data directly and is about 95\% across the entire $J/\psi$ $p_T$ region. 
The MTD trigger efficiency ($\varepsilon_{\rm trig.}$) includes the trigger electronics efficiency which varies from 95\% at low $p_T$ to more than 99\% at high $p_T$, and the online timing window cut efficiency which reaches a plateau of 99.9\%.  
The MTD efficiency ($\varepsilon_{\rm MTD}$) is determined from the cosmic ray data for the in-situ response efficiency and from the MC sample for the matching efficiency. 
It is evaluated as a function of muon $p_T$ for each MTD backleg and module separately.
Finally, the muon identification efficiency is calculated from the MC events and the plateau efficiency is above 95\% ~\cite{muid}.
Figure~\ref{fig:tot_eff} shows the individual and total efficiencies used for the $J/\psi$ cross section measurement as a function of $p_T^{J/\psi}$.

\begin{figure*}[htbp]
    \centering    
      \includegraphics[width=0.85\textwidth]{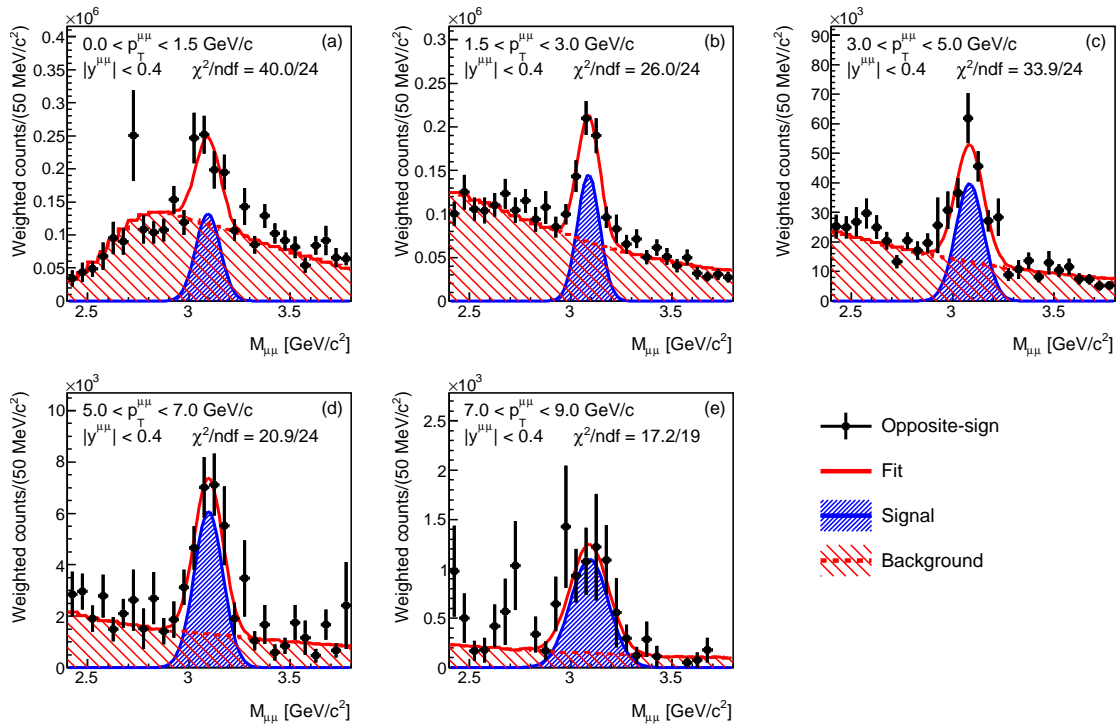}
      \caption{ Fits to the efficiency-corrected $M_{\mu\mu}$ spectra of the $J/\psi$ candidates in different $p_T$ regions. The solid red line is a combined fit to the signal and background with a single Gaussian plus a background template from the pairs of same-sign charged tracks in the TPC.   
      }
  \label{fig:jpsi_mass}
\end{figure*}

The $N_{J/\psi\to\mu^{+}\mu^{-}}^{\rm corr.}$ in different $p_T$ regions are extracted using the $\chi^2$ fit with several combinations of signal and background models of the efficiency-corrected $\mu^{+}\mu^{-}$ mass distributions. 
The signal shape is modeled by a single Gaussian, double Gaussian or Crystal-Ball function, and the background shape can be well described by the same-sign muon track pairs or a polynomial function at different orders.
The averaged result from the various fits with different shapes for signal and background is used as the mean of $N_{J/\psi}^{\rm corr.}$, and the maximum deviation from the mean is assigned as the signal extraction systematic uncertainty. 
Figure~\ref{fig:jpsi_mass} shows an example of the fit results using a single Gaussian as the signal function and same-sign muon track pairs as the background template for different $p_T$ bins. 

\subsection{Uncertainties}
\begin{figure}[!htbp]   
  \begin{center}
      \includegraphics[width=0.42\textwidth]{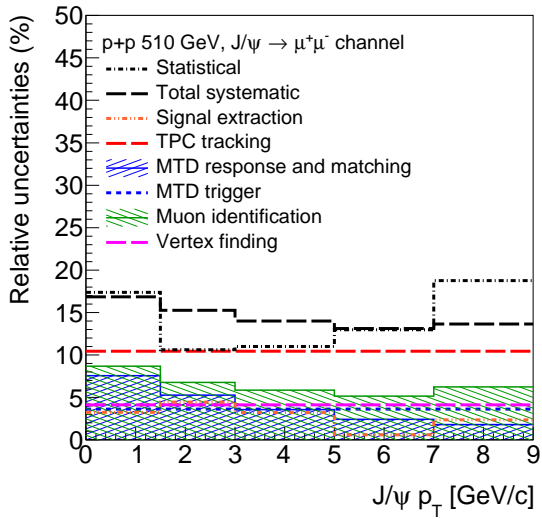}
      \end{center}
    \caption{  The statistical and individual systematic uncertainties as a function of $J/\psi$ $p_T$. The black dashed-dotted line is the statistical uncertainty; the orange dashed-dotted-dotted line is the signal extraction uncertainty; the red dashed line is the TPC tracking efficiency uncertainty; the blue shaded line is the MTD response and matching efficiency uncertainty; the blue dotted line is MTD trigger efficiency uncertainty; the green shaded line is the muon identification uncertainty; the magenta dashed line is the vertex finding efficiency uncertainty; the black dashed line is the total systematic uncertainty; A common luminosity uncertainty of 11.2\% is not included. 
  \label{fig:jpsi_xsec_sys}}
\end{figure}

The statistical uncertainty is about 10.6$-$18.8\% for different $J/\psi$ $p_T$ regions. 
There are several systematic uncertainties considered in this analysis. 
The maximum deviation in the results for variations in cuts/methods is taken as the systematic uncertainty for each source listed below.
\begin{itemize}
    \item{ The luminosity estimate and the in-bunch pile-up effect contribute global uncertainties of 8.1\%~\cite{lumi_sys} and 7.7\%, respectively. 
          The latter is estimated by comparing the results with different selections of the difference of the $z$ position measured by the VPD and TPC detectors.}
    \item{ The uncertainty in the number of $J/\psi$ candidates extraction is evaluated using different signal and background models as described in the previous section. It contributes about 0.3$-$4.8\% uncertainty depending on $J/\psi$ $p_T$. }
    \item{ The uncertainty in the TPC tracking efficiency is estimated by comparing the results using different TPC track quality selection criteria. Since it is difficult to obtain a $p_T$ dependent uncertainty due to the low statistics, the $p_T$-integrated uncertainty of 10.6\% is applied to the entire $p_T$ range.}
    \item{ The uncertainty in the MTD trigger includes two components: (i) the trigger electronics which is dominated by the statistical uncertainty in calculating the MTD trigger efficiency; (ii) the online timing window cut which is evaluated by the difference between the 2013 and 2015 data-taking. A total 3.6\% uncertainty is assigned. }   
    \item{The MTD efficiency includes three sources: (i) statistical precision of the cosmic ray data; (ii) different fit templates used for determining the response efficiency which is the main contributor; (iii) difference in the matching efficiency between cosmic ray data and simulation. The resulting uncertainty is between 1.9$-$7.6\%.}
    \item{ The uncertainty in the muon identification efficiency is determined by comparing the efficiencies from the data-driven method and the MC sample. It contributes a 5.2$-$8.7\% uncertainty depending on $J/\psi$ $p_{T}$~\cite{muid}. }
    \item{ The uncertainty in vertex finding efficiency is estimated by comparing the efficiencies from data-driven and zero-bias MC sample. A 4.1\% uncertainty is assigned.  }
\end{itemize}
The systematic uncertainties from the MTD geometric acceptance is negligible.
Systematic uncertainties from different sources are added in quadrature.
Figure~\ref{fig:jpsi_xsec_sys} shows all the uncertainties as a function of $J/\psi$ $p_T$.  

\subsection{Cross-section for $J/\psi \rightarrow \mu^{+}\mu^{-}$ }
\begin{figure}[!htbp]
  \begin{center}
      \includegraphics[width=0.42\textwidth]{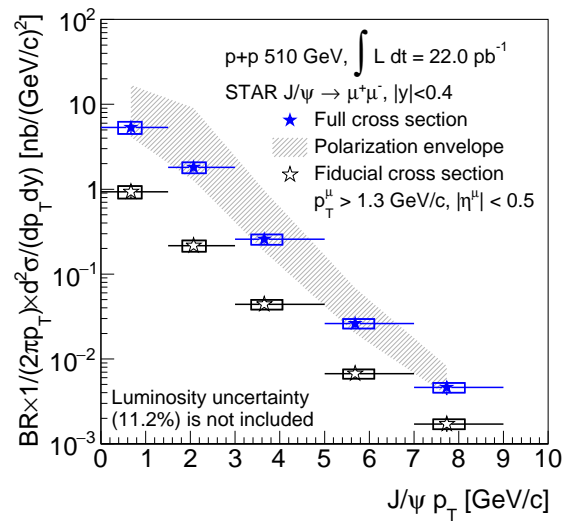}
      \end{center}
    \caption{  
     The fiducial (open black stars) and full (solid blue stars) cross sections multiplied by the branching ratio as a function of $J/\psi$ $p_T$.
     The bars and boxes are the statistical and systematic uncertainties, respectively. The gray shaded band is the polarization envelope. 
      A common luminosity uncertainty of 11.2\% is not included.
  \label{fig:xsec_dimu}}
\end{figure}

We measure the invariant differential cross section multiplied by the $\mu^{+}\mu^{-}$ branching ratio of the $J/\psi$ meson as a function of $J/\psi$ $p_T$ in a fiducial volume defined by $p_T^{\mu} > $ 1.3 GeV/$c$ and $|\eta^{\mu}| < $ 0.5 (fiducial cross section) and in a full muon decay phase space with $|y^{J/\psi}| < $ 0.4 (full cross section).
Figure~\ref{fig:xsec_dimu} shows the fiducial and full cross sections of the $J/\psi$ production. 
The fiducial cross section is calculated using the fiducial weight, $w_i^{\rm fid.} = \mathcal{A}_{\rm MTD} \times \varepsilon_{\rm reco.}$, in which the kinematic acceptance is not included.  
This eliminates the large unknown effect from the polarization assumption. 
The full cross section uses the full weight, $w_i^{\rm full} = \mathcal{A}_{J/\psi} \times \mathcal{A}_{\rm MTD} \times \varepsilon_{\rm reco.}$, to correct for the efficiency effect on each $J/\psi$ candidate. 
The central values of the full cross section in this analysis are derived under the unpolarized assumption.
The gray shaded band indicates the maximum span of the cross-sections with different polarization assumptions, denoted as the ``polarization envelope''.

Table~\ref{table:xsec_result} summarizes the results on fiducial and full cross sections of the $J/\psi$ production in different $p_T$ bins. 
The integrated fiducial and full cross sections up to 9 GeV/$c$ of $J/\psi$ $p_T$ within $|y^{J/\psi}| < $ 0.4 are 10.3 $\pm$ 0.9 (stat.) $\pm$ 1.6 (sys.) $\pm$ 1.1 (lumi.) nb and 67 $\pm$ 6 (stat.) $\pm$ 10 (sys.) $^{+200}_{-18}$ (pol.) $\pm$ 7 (lumi.) nb, respectively. 

\begin{table*}[!htbp]
  \begin{center}
   \renewcommand{\arraystretch}{1.8}
  \begin{tabular}{ccccccc}
  \hline\hline
      $p_T^{J/\psi}$ range &  & $p_T^{J/\psi}$ position &  & {\rm BR}$\times\frac{ d\sigma_{\rm fid.}^2}{2\pi p_Tdp_T dy}\pm\delta_{\rm stat.}\pm\delta_{\rm sys.}$ & & {\rm BR}$\times\frac{ d\sigma_{\rm full}^2}{2\pi p_Tdp_T dy}\pm\delta_{\rm stat.}\pm\delta_{\rm sys.}$$^{+\delta_{\rm pol.\ upper}}_{-\delta_{\rm pol.\ lower}}$ \\
      (GeV/$c$)  & & (GeV/$c$)  & & (pb/(GeV/$c$)$^2$)  & & (pb/(GeV/$c$)$^2$)   \\ 
      \hline
      0.0 - 1.5 & & 0.67 & & (9.3 $\pm$ 1.6 $\pm$ 1.5) $\times 10^{2}$ & & (5.4 $\pm$ 0.9 $\pm$ 0.9 $^{+11.5}_{-1.3}$) $\times 10^{3}$   \\
      1.5 - 3.0 & & 2.07 & & (2.2 $\pm$ 0.2 $\pm$ 0.3) $\times 10^{2}$ & & (1.8 $\pm$ 0.2 $\pm$ 0.3 $^{+7.3}_{-0.5}$) $\times 10^{3}$    \\
      3.0 - 5.0 & & 3.65 & & (4.4 $\pm$ 0.4 $\pm$ 0.6) $\times 10^{1}$ & & (2.6 $\pm$ 0.3 $\pm$ 0.4 $^{+7.2}_{-0.7}$) $\times 10^{2}$    \\
      5.0 - 7.0 & & 5.68 & & (6.7 $\pm$ 0.9 $\pm$ 0.9) $\times 10^{0}$ & & (2.6 $\pm$ 0.3 $\pm$ 0.3 $^{+4.3}_{-0.6}$) $\times 10^{1}$    \\
      7.0 - 9.0 & & 7.73 & & (1.7 $\pm$ 0.3 $\pm$ 0.2) $\times 10^{0}$ & & (4.6 $\pm$ 0.9 $\pm$ 0.6 $^{+3.7}_{-0.9}$) $\times 10^{0}$    \\
      \hline \hline
  \end{tabular}
  \end{center}
    \caption{A summary of the fiducial and full differential cross sections of the $J/\psi$ production via the $\mu^+ \mu^-$ decay channel in proton+proton collisions at $\sqrt{s} = $ 510 GeV. The fiducial volume is defined as $p_T^{\mu} > $ 1.3 GeV/$c$ and $|\eta^{\mu}| < $ 0.5. A common luminosity uncertainty of 11.2\% is not included. 
\label{table:xsec_result}}
\end{table*}


\section{Measurement of $J/\psi \rightarrow e^{+}e^{-}$ Signal}
\subsection{Data set and analysis}

The proton+proton collision data at $\sqrt{s}$ = 500 GeV, used in the $e^{+}e^{-}$ analysis, were recorded by the STAR detector in 2011.
The integrated luminosity of the data set is 22.1 $pb^{-1}$ sampled by the BEMC trigger which requires a BEMC tower with a transverse energy deposit larger than 4.3 GeV~\cite{jpsi_200}.
The $e^{\pm}$ candidates are reconstructed and identified using information from the TPC and BEMC detectors. 
The track quality requirements are that each track has at least 25 out of 45 possible hits in the TPC, the number of hits for the $dE/dx$ measurement must be larger than 15 to ensure a good $dE/dx$ resolution, and tracks are reconstructed within the TPC acceptance of $|\eta|<$ 1. 
The electron and positron candidates are then identified by their ionization energy loss ($\langle dE/dx \rangle$) in the TPC. 
The normalized $\langle dE/dx \rangle$ is defined as follows:
\begin{eqnarray}
    n\sigma_{e} = \frac{1}{\sigma_{dE/dx}}\ln(\frac{\langle dE/dx \rangle^{\rm me.}}{\langle dE/dx \rangle^{\rm exp.}_{e}}), 
\label{eq:nsigmaE}
\end{eqnarray}
where $\langle dE/dx \rangle^{\rm me.}$ and $\langle dE/dx \rangle^{\rm exp.}_{e}$ are the measured $\langle dE/dx \rangle$ and the expected $\langle dE/dx \rangle$ value for electron, and the $\sigma_{dE/dx}$ is the experimental $\ln(dE/dx)$ resolution.
The $n\sigma_{e}$ requirement for the triggered $e^{\pm}$ candidates is set to be $|n\sigma_{e}| <$ 2.  
The triggered $e^{\pm}$ candidate is also required to have $p_T >$ 3.5 GeV/$c$, its track must have DCA from the primary vertex less than 1 cm to reduce contamination from pile-up tracks, and it is required to match to a BEMC trigger tower in which the ADC value is larger than 290, corresponding to a deposited energy of 4.3 GeV.  
A cut on the ratio of the momentum measured by the TPC to the energy deposited in the BEMC towers, 0.3 $< pc/E <$ 1.5, is used to further suppress contribution of hadrons in triggered electron selection. 
However, the non-triggered $e^{\pm}$ candidate is only required to be a TPC track with $p_T >$ 1 GeV/$c$ and DCA $<$ 3 cm.
The looser DCA requirement is applied to increase the statistics, since lower $p_T$ tracks are more affected by multiple scattering. 

\begin{figure}[h]
  \begin{center}
      \includegraphics[width=0.4\textwidth]{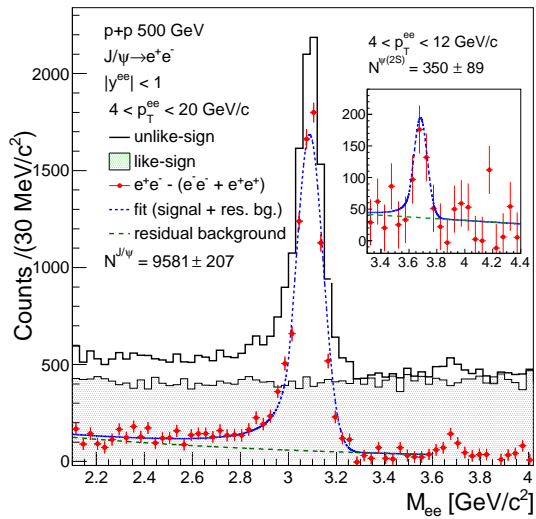}
      \end{center}
    \caption{ The invariant mass distributions of $e^{+}e^{-}$ pairs before and after the like-sign background (the green histogram) subtraction, as shown in black histogram and red solid circles, respectively.  
    The blue curve is a fit to the mass spectrum. The green-dashed line indicates the residual background and a Crystal-Ball function is used to describe the $J/\psi$ and $\psi(2S)$ signals, where $\psi(2S)$ is shown in the insert. The error bars depict the statistical uncertainties.}
  \label{fig:jpsi_mass_diele}
\end{figure}

\begin{figure*}[!htbp]
    \centering    
      \includegraphics[width=1.0\textwidth]{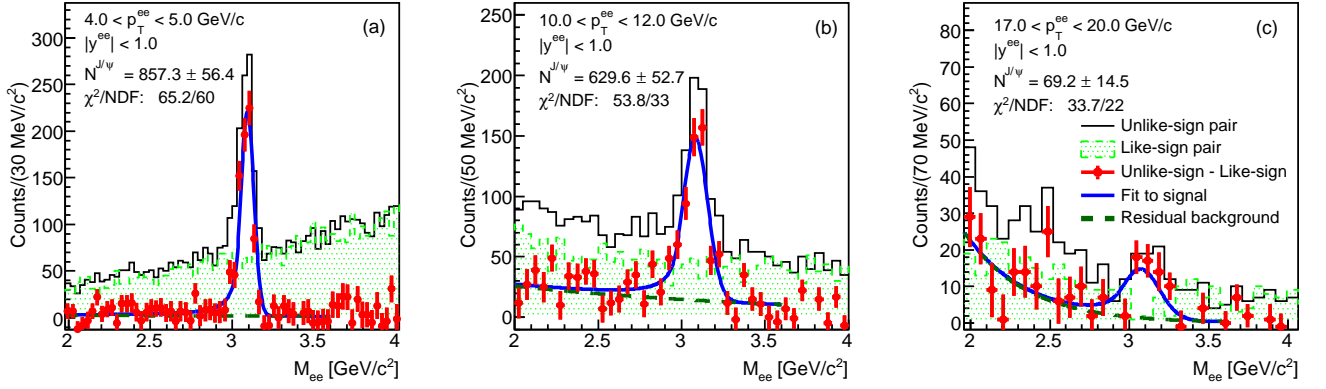}
      \caption{ Dielectron mass distributions after the like-sign background subtraction (red solid circles) in different $p_T$ ranges. 
      The solid blue line is a combined fit to the signal and residual background with a Crystal-Ball function plus an exponential function. 
      The error bars depict the statistical error.
      }
  \label{fig:jpsi_ee_signals}
\end{figure*}

Figure~\ref{fig:jpsi_mass_diele} shows both $J/\psi$ and $\psi(2S)$ signals which are reconstructed from $e^{+}e^{-}$ pairs, which include the triggered electron pairing with either another triggered electron or with a non-triggered electron.
The signal shape for $J/\psi$ candidates is obtained from the MC simulation, which includes track momentum resolution and electron bremsstrahlung radiation in the detector.
On the other hand, the background includes the combinatorial background evaluated using the like-sign $e^{+}e^{+}$ and $e^{-}e^{-}$ pairs within the same event (green histogram) and the residual background, which mainly comes from the Drell-Yan process, $c\bar{c}$ and $b\bar{b}$ decays. 
The invariant mass distribution after the like-sign background subtraction is fitted with a $J/\psi$ signal shape combined with an exponential function. 
The raw $J/\psi$ yield is obtained by counting the bin contents after subtracting the residual background in the mass range of 2.7 $< M_{ee} <$ 3.3 GeV/$c^{2}$. 
There are 9581 $\pm$ 207 $J/\psi$ signals in 4 $< p_T^{J/\psi} <$ 20 GeV/$c$.   
About $\sim$10$\%$ of $J/\psi$ candidates are reconstructed outside this mass window based on MC simulations, and  the $J/\psi$ raw yield as a function of $p_T$ is corrected for this effect. 
A total of 350 $\pm$ 89 $\psi(2S)$ signals are obtained in the mass counting range of 3.5 $< M_{ee} <$ 3.8 GeV/$c^{2}$. 
The $\psi(2S)$ signals are extracted by using the same method as for $J/\psi$, but with a linear function to describe the residual background.
Figure~\ref{fig:jpsi_ee_signals} shows invariant mass distributions of $e^{+}e^{-}$ pairs (black histogram) and the like-sign (filled histogram) $e^{+}e^{+}$ and $e^{-}e^{-}$ pairs pairs in three representative $p_{T}$ bins. 

The measurement of the $J/\psi$ differential production cross section multiplied by BR for the $e^{+}e^{-}$ decay channel, ($5.971 \pm 0.032$)\%~\cite{pdg_jpsi}, is defined as:
\begin{eqnarray}
    {\rm BR}\times\frac{d^2\sigma}{2\pi p_T dp_T dy} = \frac{N^{\rm raw}_{J/\psi \to e^{+}e^{-}}}{(2\pi p_T) \cdot \int \mathcal{L} dt \cdot \mathcal{A}\varepsilon  \cdot \Delta p_T \cdot \Delta y}, 
\label{eq:xsec_ee}
\end{eqnarray}
where ${\rm BR}$ is the branching ratio for the $J/\psi \to e^+e^-$ decay channel; $N^{\rm raw}_{J/\psi \to e^{+}e^{-}}$ is the raw number of reconstructed $J/\psi$ via the $e^{+}e^{-}$ pairs; $\mathcal{A}\varepsilon$ is the detector's geometric acceptance times the detection efficiency of the $J/\psi$ candidates; $\int \mathcal{L} dt$, $\Delta p_T$, and $\Delta y$ have the same meanings as of that in the $\mu^{+}\mu^{-}$ decay channel.

\begin{figure}[!htbp]
  \begin{center}
      \includegraphics[width=0.42\textwidth]{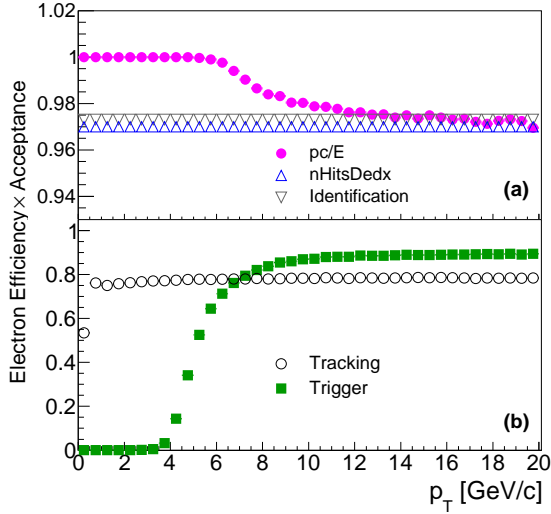}
      \end{center}
    \caption{ The individual electron efficiencies for the triggered electrons as a function of $p_T^e$ including $pc/E$ cut (magenta solid circles), the number of hits for the $dE/dx$ measurement (blue up-triangles), identification (gray down-triangles), tracking (black open circles), and trigger (green solid boxes).} 
  \label{fig:eff_ele}
\end{figure}

The total $J/\psi$ detection efficiency, $\mathcal{A}\varepsilon$, includes the detector acceptance, the mass bin counting efficiency, and individual efficiency of the electron candidates including the TPC tracking efficiency, the electron identification, the selection on the number of hits for the $dE/dx$ measurement, the additional efficiency for the trigger, and the cut on $pc/E$ for the triggered electrons. The decay electron's momentum resolution and additional $p_{T}$ smearing are also included in the calculation of the $J/\psi$ detection efficiency.
The efficiencies for the number of $dE/dx$ hits and $n\sigma_{e}$ cuts are assessed using a pure electron sample from photon conversion data, while all the other acceptance and efficiencies are obtained from MC simulation with the STAR detector geometry. 
Figure~\ref{fig:eff_ele} shows the individual efficiencies for the triggered electron candidates as a function of $p_{T}^{e}$.

\begin{figure}[!htbp]
  \begin{center}
      \includegraphics[width=0.4\textwidth]{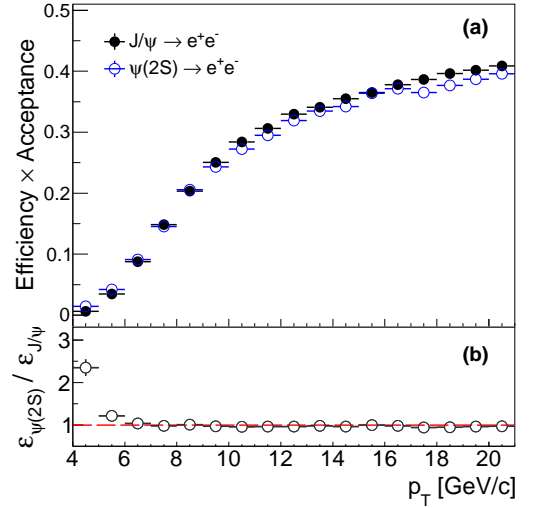}
      \end{center}
  \caption{(a) The $J/\psi$ and $\psi(2S)$ detection efficiencies as a function of $p_T$ as shown in the black solid circles and the blue open circles, respectively. (b) The detection efficiency ratio of $\psi(2S)$ to $J/\psi$  as a function of $p_T$ and the red dashed-line is a line at unity.} 
  \label{fig:jpsi_psi2s_eff}
\end{figure}

The detection efficiency for $\psi(2S)$ candidates is obtained in the same way as for $J/\psi$.  
The relatively larger invariant mass of $\psi(2S)$ enhances the trigger efficiency in the low $p_{T}$ range while
the slightly larger opening angle between the electron and positron daughters decaying from the $\psi(2S)$ will result in a smaller acceptance and thus a lower detection efficiency at high $p_{T}$. 
Figure~\ref{fig:jpsi_psi2s_eff} shows the detection efficiency for $J/\psi$ and $\psi(2S)$, as well as the $\psi(2S)$ to $J/\psi$ efficiency ratio as a function of $p_T^{e^{+}e^{-}}$.  

\subsection{Uncertainties}
\begin{figure}[!htbp]
  \begin{center}
      \includegraphics[width=0.44\textwidth]{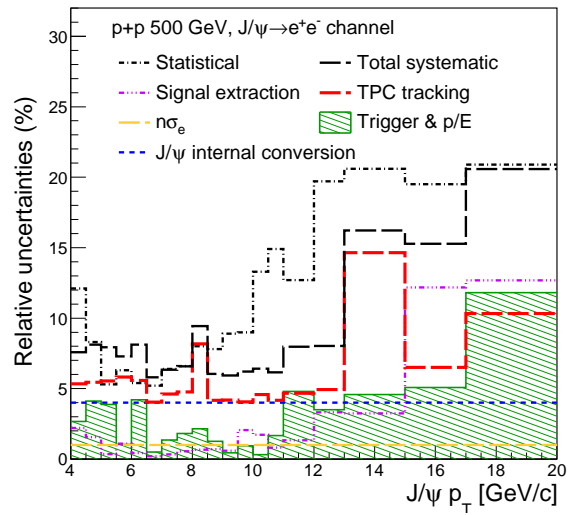}
      \end{center}
    \caption{The statistical and individual systematic uncertainties as a function of $J/\psi$ $p_{T}$. The black dashed-dotted line is the statistical uncertainty. The violet dashed-dotted-dotted line is the signal extraction uncertainty; the red dashed line is the TPC tracking uncertainty; the yellow dashed line is the electron identification ($n\sigma_{e}$) uncertainty; the green shaded line is the trigger uncertainty; the blue dashed line is the $J/\psi$ internal conversion uncertainty; the black dashed line is the total systematic uncertainty;  A common luminosity uncertainty of 8.1\% is not included.} 
  \label{fig:Unsys}
\end{figure}

The systematic uncertainties for the final $J/\psi$ cross section are estimated by varying analysis selections in both data and MC simulation and comparing the corresponding $J/\psi$ cross section to the nominal value.
The systematic uncertainties considered in this analysis are the following:
\begin{itemize}
    \item{The uncertainty in the luminosity contributes an overall 8.1\% uncertainty~\cite{lumi_sys}. The uncertainty in the in-bunch pile-up effect is negligible due to the low instantaneous luminosity in the 2011 data-taking and the high vertex finding efficiency for the BEMC triggered events~\cite{pile_up}.}
	\item{The $J/\psi$ extraction uncertainty is estimated by using different fitting ranges and different residual background shapes. It contributes a 0.2$-$12.7\% uncertainty depending on $J/\psi$ $p_{T}$.}
	\item{The uncertainty in the TPC tracking efficiency is estimated in the same way as in the $\mu^{+}\mu^{-}$ decay channel, and it contributes a 4$-$14\%. } 
    \item{The uncertainty in the trigger efficiency is evaluated by comparing BEMC response in data and MC simulation, and the contribution is a 0.3$-$11.8\% in various $J/\psi$ $p_{T}$ range.} 
	\item{The uncertainty in the electron identification is estimated by comparing the difference between photonic electron's $n\sigma_{e}$ distribution at different $p_{T}$ ranges, and it contributes an overall 1\% uncertainty.}
	\item{The $J/\psi$ internal conversion is estimated to be 4\% within the mass counting range of 2.7 to 3.3 GeV/$c^2$.}
\end{itemize}
The systematic uncertainty from the vertex finding is negligible. 
Systematic uncertainties from different sources are added in quadrature. 
Figure~\ref{fig:Unsys} shows all the uncertainties as a function of $J/\psi$ $p_T$.
For the $\psi(2S)$ to $J/\psi$ ratio, most of the systematic uncertainties cancel in the ratio, except for the signal extraction and trigger efficiency. 
They are evaluated the same way as for the $J/\psi$. 
They contribute 5.5\% and 5.3\%, respectively, to the uncertainty of the ratio measurement. 

\subsection{Cross-section for $J/\psi \to e^+ e^-$}
\begin{figure}[!htbp]
  \begin{center}
      \includegraphics[width=0.44\textwidth]{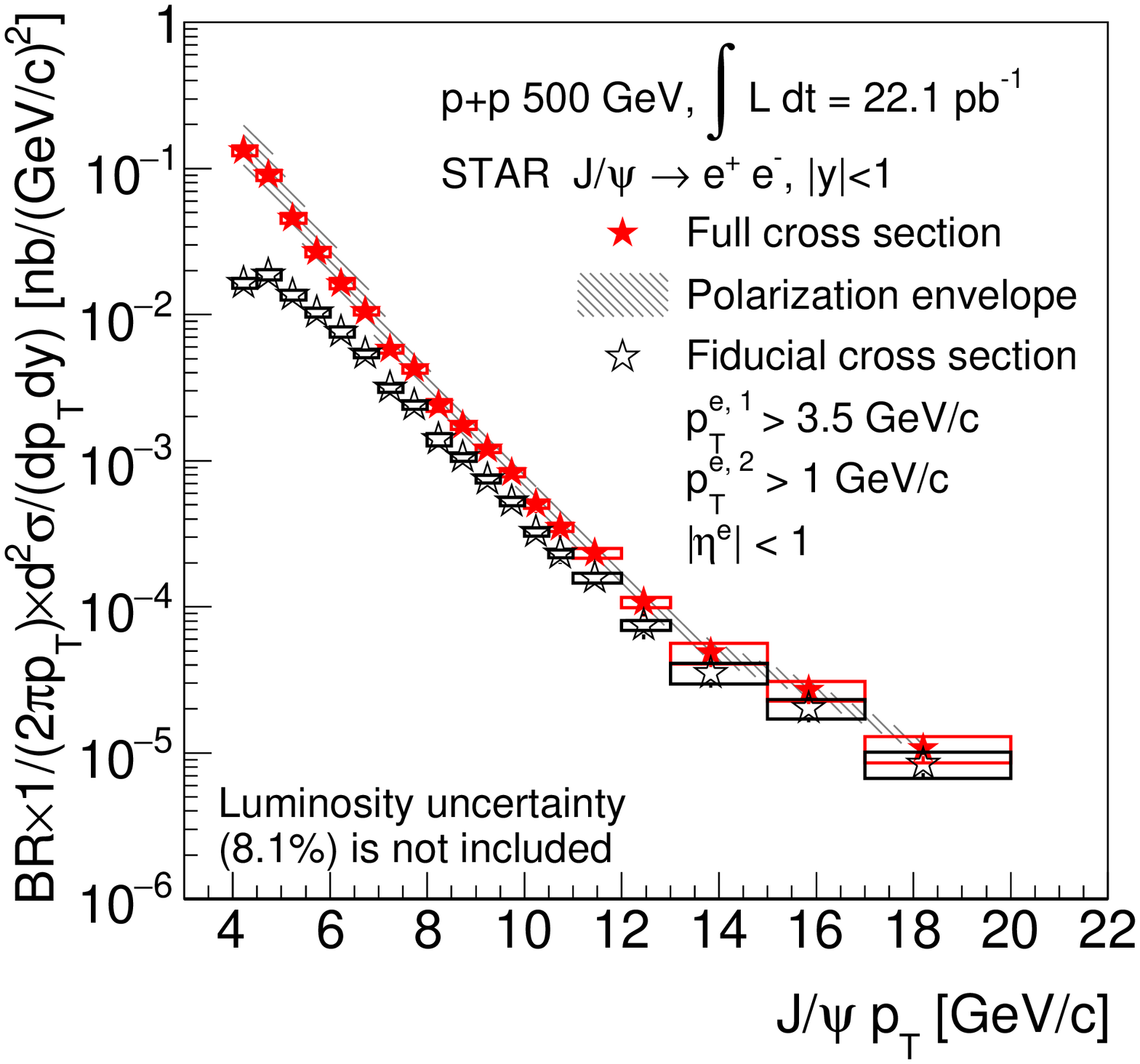}
      \end{center}
  \caption{  
      The fiducial (open black stars) and full (solid red stars) cross sections multiplied by the branching ratio as a function of $J/\psi$ $p_T$. 
      The boxes are the total systematic uncertainty. The bars are the statistical uncertainty and are too small to be visible in the figure.
      The gray shaded band is the polarization envelope. A common luminosity uncertainty of 8.1\% is not included.
  \label{fig:xsec_ee}}
\end{figure}

Figure~\ref{fig:xsec_ee} shows the fiducial and full production cross sections measured in the $e^{+}e^{-}$ decay channel for 4 $< p_T^{J/\psi} <$ 20 GeV/$c$ and $|y| < $ 1.0. 
Table~\ref{table:xsec_result_diele} summarizes the cross sections of the $J/\psi$ production as a function of $p_T$. 
Similarly as for the $\mu^{+}\mu^{-}$ channel, the full cross section is calculated under the unpolarized assumption and the polarization envelope is obtained from the same five extreme cases. 
This provides us the information of the $J/\psi$ production cross section within the full $J/\psi$ decay phase space. 
On the other hand, the fiducial cross section only accesses the restricted phase space, but it is independent from any polarization assumptions.  
The integrated fiducial and full cross sections from 4 to 20 GeV/$c$ of $J/\psi$ $p_T$ are 2.90 $\pm$ 0.08 (stat.) $\pm$ 0.22 (sys.) $\pm$ 0.24 (lumi.) nb and 10.7 $\pm$ 0.5 (stat.) $\pm$ 0.8 (sys.) $^{+5.9}_{-2.2}$ (pol.) $\pm$ 0.9 (lumi.) nb, respectively.

\begin{table*}[!htbp]
  \begin{center}    
     \renewcommand{\arraystretch}{1.8}
  \begin{tabular}{ccccccc}    
  \hline\hline
$p_T^{J/\psi}$ range &  & $p_T^{J/\psi}$ position &  & {\rm BR}$\times\frac{d\sigma_{\rm fid.}^2}{2\pi p_Tdp_T dy}\pm\delta_{\rm stat.}\pm\delta_{\rm sys.}$ & & {\rm BR}$\times\frac{d\sigma_{\rm full}^2}{2\pi p_Tdp_T dy}\pm\delta_{\rm stat.}\pm\delta_{\rm sys.}$$^{+\delta_{\rm pol.\ upper}}_{-\delta_{\rm pol.\ lower}}$ \\
      (GeV/$c$)  & & (GeV/$c$)  & & (pb/(GeV/$c$)$^2$)  & & (pb/(GeV/$c$)$^2$)   \\ 
      \hline
      4.0 - 4.5 & & 4.23 & & (1.64 $\pm$ 0.20 $\pm$ 0.12) $\times 10^{1}$ & & (1.32 $\pm$0.16 $\pm$ 0.10 $^{+0.72}_{-0.35}$) $\times 10^{2}$   \\
      4.5 - 5.0 & & 4.73 & & (1.88 $\pm$ 0.16 $\pm$ 0.15) $\times 10^{1}$ & & (9.0 $\pm$0.8 $\pm$ 0.7 $^{+5.2}_{-2.0}$) $\times 10^{1}$   \\
      5.0 - 5.5 & & 5.23 & & (1.36 $\pm$ 0.07 $\pm$ 0.11) $\times 10^{1}$ & & (4.6 $\pm$0.2 $\pm$ 0.4 $^{+2.6}_{-0.8}$) $\times 10^{1}$   \\
      5.5 - 6.0 & & 5.73 & & (10.3 $\pm$ 0.7 $\pm$ 0.8) $\times 10^{0}$ & & (2.69 $\pm$0.17 $\pm$ 0.20 $^{+1.59}_{-0.39}$) $\times 10^{1}$   \\
      6.0 - 6.5 & & 6.23 & & (7.6  $\pm$ 0.4 $\pm$ 0.6) $\times 10^{0}$ & & (1.64 $\pm$0.09 $\pm$ 0.13 $^{+0.97}_{-0.24}$) $\times 10^{1}$   \\
      6.5 - 7.0 & & 6.73 & & (5.40 $\pm$ 0.29 $\pm$ 0.31) $\times 10^{0}$ & & (10.6 $\pm$0.6 $\pm$ 0.6 $^{+5.4}_{-1.6}$) $\times 10^{0}$   \\
      7.0 - 7.5 & & 7.23 & & (3.15 $\pm$ 0.20 $\pm$ 0.20) $\times 10^{0}$ & & (5.8 $\pm$0.4 $\pm$ 0.4 $^{+2.9}_{-0.9}$) $\times 10^{0}$   \\
      7.5 - 8.0 & & 7.73 & & (2.41 $\pm$ 0.16 $\pm$ 0.16) $\times 10^{0}$ & & (4.26 $\pm$0.28 $\pm$ 0.28 $^{+1.76}_{-0.61}$) $\times 10^{0}$   \\
      8.0 - 8.5 & & 8.23 & & (1.40 $\pm$ 0.11 $\pm$ 0.13) $\times 10^{0}$ & & (2.40 $\pm$0.19 $\pm$ 0.23 $^{+0.97}_{-0.33}$) $\times 10^{0}$   \\
      8.5 - 9.0 & & 8.73 & & (10.6 $\pm$ 0.8 $\pm$ 0.6) $\times 10^{-1}$ & & (1.75 $\pm$0.14 $\pm$ 0.11 $^{+0.65}_{-0.23}$) $\times 10^{0}$   \\
      9.0 - 9.5 & & 9.24 & & (7.5 $\pm$ 0.7 $\pm$ 0.4) $\times 10^{-1}$ & & (12.0 $\pm$1.1 $\pm$ 0.7 $^{+4.0}_{-1.6}$) $\times 10^{-1}$   \\
      9.5 - 10.0 & & 9.73 & & (5.26 $\pm$ 0.48 $\pm$ 0.33) $\times 10^{-1}$ & & (8.3 $\pm$0.7 $\pm$ 0.5 $^{+2.4}_{-1.1}$) $\times 10^{-1}$   \\
      10.0 - 10.5 & & 10.23 & & (3.26 $\pm$ 0.43 $\pm$ 0.21) $\times 10^{-1}$ & & (5.06 $\pm$0.67 $\pm$ 0.32 $^{+1.57}_{-0.62}$) $\times 10^{-1}$   \\
      10.5 - 11.0 & & 10.74 & & (2.31 $\pm$ 0.35 $\pm$ 0.14) $\times 10^{-1}$ & & (3.51 $\pm$0.52 $\pm$ 0.22 $^{+1.02}_{-0.42}$) $\times 10^{-1}$   \\
      11.0 - 12.0 & & 11.44 & & (1.58 $\pm$ 0.20 $\pm$ 0.13) $\times 10^{-1}$ & & (2.33 $\pm$0.30 $\pm$ 0.19 $^{+0.62}_{-0.27}$) $\times 10^{-1}$   \\
      12.0 - 13.0 & & 12.45 & & (7.5 $\pm$ 1.5 $\pm$ 0.6) $\times 10^{-2}$ & & (10.7 $\pm$2.1 $\pm$ 0.9 $^{+2.3}_{-1.2}$) $\times 10^{-2}$   \\
      13.0 - 15.0 & & 13.83 & & (3.5 $\pm$ 0.7 $\pm$ 0.6) $\times 10^{-2}$ & & (4.9 $\pm$1.0 $\pm$ 0.8 $^{+1.0}_{-0.5}$) $\times 10^{-2}$   \\
      15.0 - 17.0 & & 15.85 & & (2.02 $\pm$ 0.39 $\pm$ 0.31) $\times 10^{-2}$ & & (2.7 $\pm$0.5 $\pm$ 0.4 $^{+0.4}_{-0.3}$) $\times 10^{-2}$   \\
      17.0 - 20.0 & & 18.20 & & (0.84 $\pm$ 0.18 $\pm$ 0.17) $\times 10^{-2}$ & & (1.07 $\pm$0.23 $\pm$ 0.22 $^{+0.13}_{-0.09}$) $\times 10^{-2}$   \\
      \hline \hline
 \end{tabular}
  \end{center}      
 \caption{A summary of the fiducial and full differential cross sections for the inclusive $J/\psi$ production via the $e^+ e^-$ decay channel in proton+proton collisions at $\sqrt{s} = $ 500 GeV. A common luminosity uncertainty of 8.1\% is not included.
\label{table:xsec_result_diele}}
\end{table*}

\begin{figure}[!htbp]
  \begin{center}
      \includegraphics[width=0.44\textwidth]{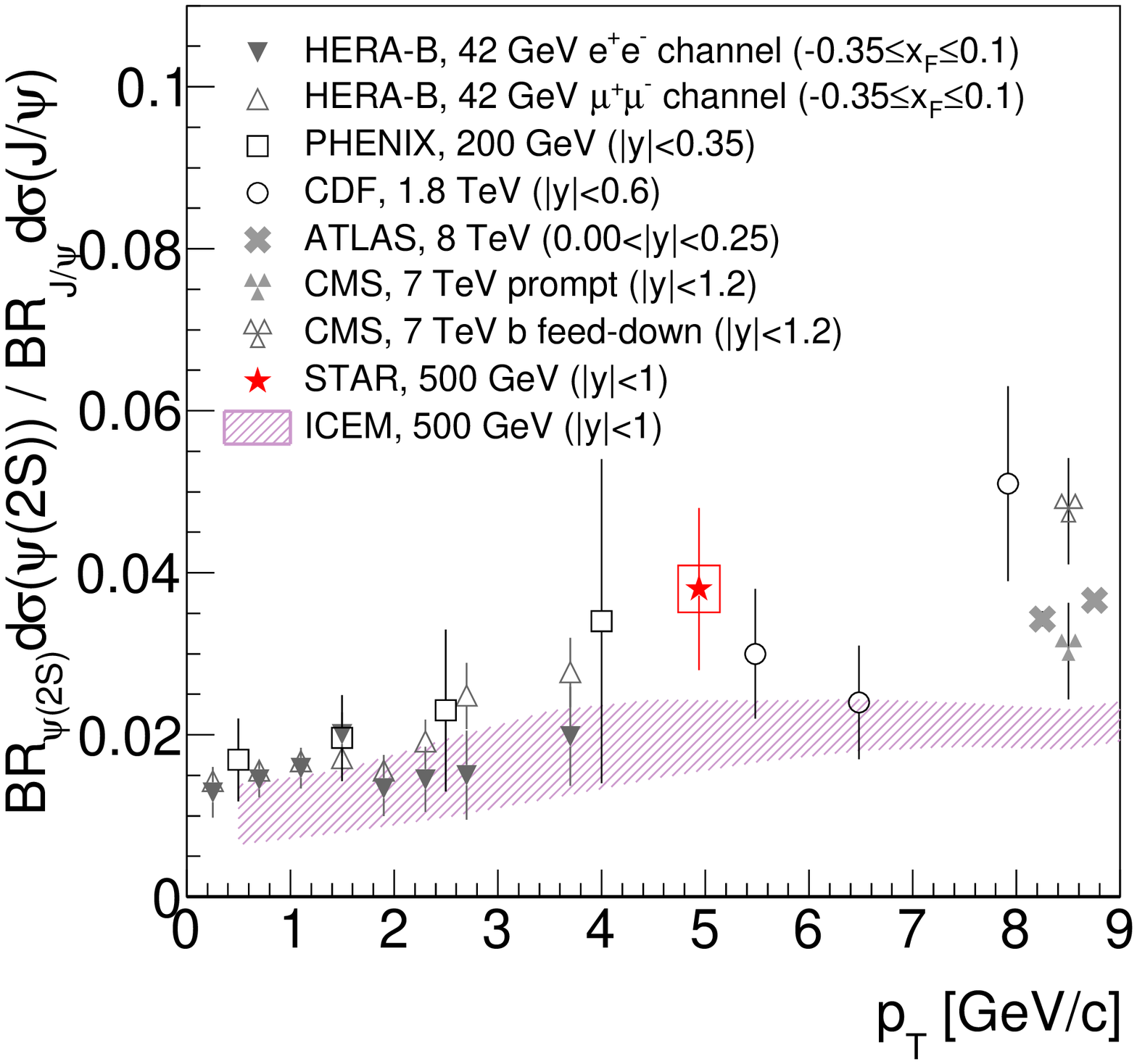}
      \end{center}
    \caption{  The cross section ratio of $\psi(2S)$ over $J/\psi$ as a function of their $p_T$ measured by STAR (red solid star) is compared to results from CDF (open circles), PHENIX (open boxes), HERA-B (solid-down-triangles and open-up-triangles), ATLAS (gray crosses), CMS (gray solid and open triangle-crosses) experiments, and the prediction from ICEM (purple band). The bar and box indicates the statistical and total systematic uncertainty, respectively.
  \label{fig:psi2S}}
\end{figure}

The cross section ratio of $\psi(2S)$ over $J/\psi$ is 0.038 $\pm$ 0.010 (stat.) $\pm$ 0.003 (sys.) measured in the $p_T$ range of 4 $< p_T^{\rm meson} <$ 12 GeV/$c$, which is shown in Fig.~\ref{fig:psi2S}. 
The result is consistent with other experimental measurements~\cite{cdf_psi2s, phenix_psi2s, hera_b_psi2s, atlas_psi2s, cms_psi2s} and there is no obvious dependence on the collision energy observed.
The ICEM model~\cite{icem} prediction is consistent with our data within uncertainties.


\section{Combined results}

\begin{figure}[!htbp]
  \begin{center}
      \includegraphics[width=0.48\textwidth]{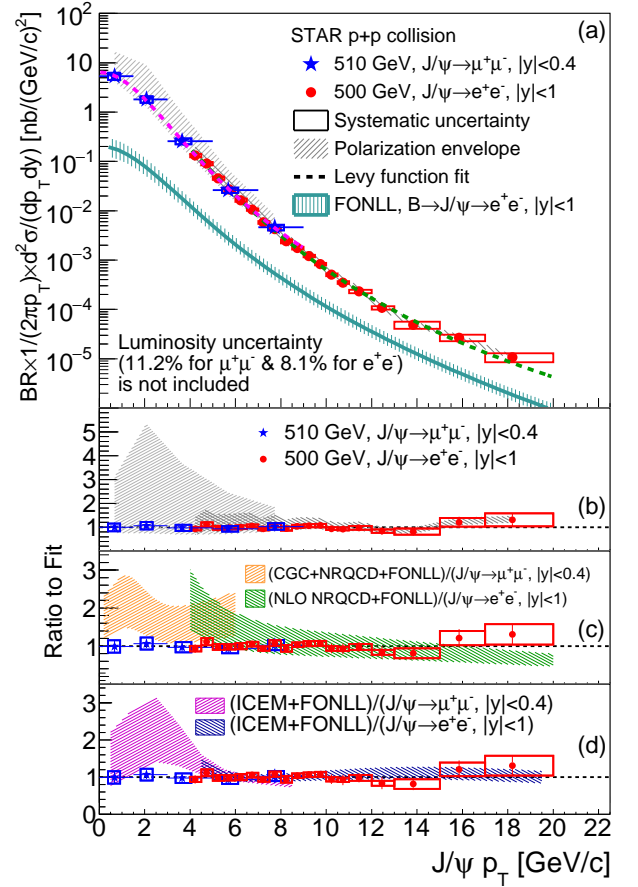}
      \end{center}
    \caption{  (a) The $J/\psi$ differential full production cross sections as a function of $p_T^{J/\psi}$ in proton+proton collisions at $\sqrt{s}$ = 510 and 500 GeV measured through the $\mu^{+}\mu^{-}$ (blue stars) and $e^{+}e^{-}$ decay channels (red circles). The shaded region around the data points denotes the polarization envelope and the green curve is the estimation of the $B$-hadrons feed-down from FONLL~\cite{fonll}.
    (b, c, d) Ratios of data and different model calculations to the Levy fit function.
    The bars and boxes indicate the statistical and total systematic uncertainty, respectively. 
    A common luminosity uncertainties of 11.2\% and 8.1\% for the $\mu^{+}\mu^{-}$ and $e^{+}e^{-}$ decay channels are not included.
  \label{fig:xsec_tot}}
\end{figure}

Figure~\ref{fig:xsec_tot} shows the differential cross section of inclusive $J/\psi$ production in proton+proton collisions at $\sqrt{s}$ = 510 and 500 GeV measured by the STAR experiment combining the $\mu^{+}\mu^{-}$ and $e^{+}e^{-}$ decay channels. 
Please note that there is a $\sim$3\% difference between the cross sections at 510 and 500 GeV collision energies and the difference between the different rapidity coverages is negligible~\cite{pythia8}. 
The unpolarized-assumption result is compared to the NRQCD~\cite{nrqcd} and ICEM~\cite{icem} calculations of $J/\psi$ production, which includes feed-down contributions from excited charmonium states.
The prediction from the CGC effective theory coupled with NRQCD (CGC+NRQCD)~\cite{cgc} lies systematically above the data at low $p_T$, however, it is consistent with the data within the polarization envelope.  
The NLO NRQCD~\cite{nrqcd} result does a reasonably good job in describing the data above 6 GeV/$c$. 
The ICEM calculation~\cite{icem} can cover the entire $p_T$ range and is also consistent with the data within the polarization envelope. 
The feed-down contributions from $B$-hadrons are about 10$-$20\% in the $p_T < $ 10 GeV/$c$ region and nearly 40\% in our maximum $p_T$ bin (20 GeV/$c$) as measured by other experiments~\cite{jpsi_atlas, jpsi_cms}.
Therefore, to present a fair comparison, all of the predictions are adjusted to include this contribution using the FONLL calculation~\cite{fonll}, shown as a green band in Fig.~\ref{fig:xsec_tot} (a). 

\begin{figure}[!htbp]
  \begin{center}
      \includegraphics[width=0.4\textwidth]{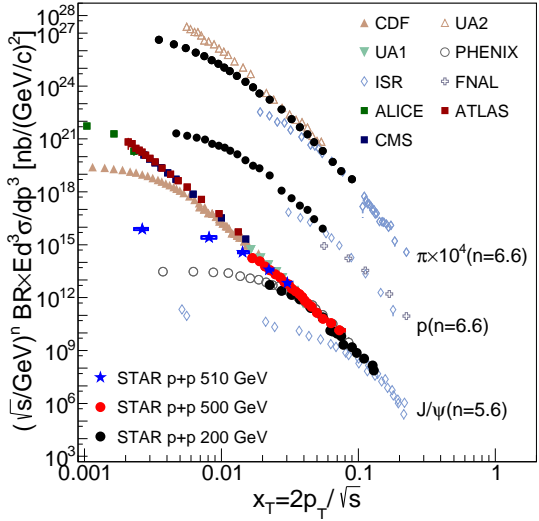}
      \end{center}
  \caption{ The $x_T$ dependence of pion, proton, and $J/\psi$ from different experiments.  
  \label{fig:xT}}
\end{figure}

The scaling behavior of particle production with $x_T = 2p_T/\sqrt{s}$ is characteristic for production through fragmentation due to hard scatterings. 
The $x_T$ scaling ($E\frac{d^3\sigma}{dp^3} = g(x_T)/s^{n/2}$) has been tested for pions, protons, and the $J/\psi$ for various collision energies~\cite{xt_first}, where $n$ is a free parameter which can be interpreted as the number of active partons involved in hadron production. 
Figure~\ref{fig:xT} shows the $x_T$ dependence of protons, pions, and $J/\psi$.  
The $J/\psi$ measured in 510 and 500 GeV proton+proton collisions has been fit to extract the parameter $n$, with $n=5.6 \pm 0.1$. 
This value is consistent with $n=5.6 \pm 0.2$ as found in a previous STAR measurement~\cite{star_xt}, as well as other previous measurements ~\cite{xt_results_cdf, cdf_psi2s, xt_results_ua1, xt_results_phenix, xt_results_isr, xt_results_star, xt_results_star_2, xt_results_8, xt_results_9, xt_results_10} at high-$p_T$, and this value is close to the CO and CEM predictions, which are $n\sim6$~\cite{xt_pred_1, xt_pred_2} and smaller than that from NNLO$^{\star}$ CSM prediction which is $n\sim8$~\cite{xt_pred_3}. 
The broken scaling at low-$p_T$ is due to the onset of soft processes~\cite{star_xt}.

\section{Conclusions}
Differential cross sections for the $J/\psi$ meson in proton+proton collisions at $\sqrt{s} = $ 510 and 500 GeV at RHIC are measured using the $\mu^{+}\mu^{-}$ and $e^{+}e^{-}$ decay channels. 
The results cover a wide $p_T^{J/\psi}$ range from 0 to 20 GeV/$c$ within $|y^{J/\psi}| < $ 0.4 and 1.0 for the $\mu^{+}\mu^{-}$ and $e^{+}e^{-}$ channel, respectively. 
Two different measurements of the inclusive $J/\psi$ production cross section have been presented. 
The first is a fiducial cross section measurement utilizing only a restricted phase space as defined by detector acceptance. 
It is independent of the assumptions regarding polarization which results in a large systematic uncertainty at low $p_T$.
This allows direct comparisons between measurements and theoretical calculations in the future, for more discriminating tests of the models.
The second is a full cross section measurement, accessing the full $J/\psi$ decay phase space, depending highly on assumptions regarding polarization. 
The integrated fiducial and full production cross sections measured for inclusive $J/\psi$ mesons within $0 < p_T^{J/\psi} < 9$ GeV/$c$ are 10.3 $\pm$ 0.9 (stat.) $\pm$ 1.6 (sys.) $\pm$ 1.1 (lumi.) nb and 67 $\pm$ 6 (stat.) $\pm$ 10 (sys.) $^{+200}_{-18}$ (pol.) $\pm$ 7 (lumi.) nb, respectively, via the $\mu^{+}\mu^{-}$ channel. 
For $4 < p_T^{J/\psi} < 20$ GeV/$c$, they are 2.90 $\pm$ 0.08 (stat.) $\pm$ 0.22 (sys.) $\pm$ 0.24 (lumi.) nb and 10.7 $\pm$ 0.5 (stat.) $\pm$ 0.8 (sys.) $^{+5.9}_{-2.2}$ (pol.) $\pm$ 0.9 (lumi.) nb, respectively, via the $e^{+}e^{-}$ channel. 
The calculations from CGC+NRQCD, NLO NRQCD and ICEM~\cite{cgc, nrqcd}, which cover low, high, and both $p_T$ regions respectively, give a reasonable description for the data within the polarization envelope.  
The $x_T$ dependence for $J/\psi$ production is also presented for 510 and 500 GeV proton+proton collisions. 
The result is consistent with measurements at other collision energies from other collaborations.  
The ratio of $\psi(2S)$ to $J/\psi$ for $p_T$ from 4$-$12 GeV/$c$ is measured to be 0.038 $\pm$ 0.010 (stat.) $\pm$ 0.003 (sys.). 
It is consistent with results from other experiments and there is no obvious collision energy dependence.
Since the $J/\psi$ production mechanism is not yet fully understood, it is important to continue confronting the models that incorporate the most current understanding with new data. 
A more discriminating comparison to theoretical models at low $p_T$ can be performed in the future, if the calculations are carried out within the fiducial volume of the STAR detector, eliminating the uncertainty due to the $J/\psi$ polarization.
The results presented in this paper, together with cross section measurements at other energies, and measurements of the polarization, contribute to the goal of better understanding the production of heavy quarkonium in hadronic collisions.


\section{Acknowledgments}
We thank the RHIC Operations Group and RCF at BNL, the NERSC Center at LBNL, and the Open Science Grid consortium for providing resources and support.  This work was supported in part by the Office of Nuclear Physics within the U.S. DOE Office of Science, the U.S. National Science Foundation, the Ministry of Education and Science of the Russian Federation, National Natural Science Foundation of China, Chinese Academy of Science, the Ministry of Science and Technology of China and the Chinese Ministry of Education, the National Research Foundation of Korea, Czech Science Foundation and Ministry of Education, Youth and Sports of the Czech Republic, Hungarian National Research, Development and Innovation Office, New National Excellency Programme of the Hungarian Ministry of Human Capacities, Department of Atomic Energy and Department of Science and Technology of the Government of India, the National Science Centre of Poland, the Ministry  of Science, Education and Sports of the Republic of Croatia, RosAtom of Russia and German Bundesministerium fur Bildung, Wissenschaft, Forschung and Technologie (BMBF) and the Helmholtz Association.


\end{document}